# Scattering from dilute ferrofluid suspensions in soft polymer gels

Alvaro V. N. C. Teixeira,[1,2,*] Isabelle Morfin,[1] Françoise Ehrburger-Dolle,[1] Cyrille Rochas,[1] Erik Geissler,[1] Pedro Licinio,[2] and Pierre Panine[3]

[1] *Laboratoire de Spectrométrie Physique UMR CNRS 5588, Université Joseph Fourier de Grenoble, Boîte Postale 87, 38402 St. Martin d'Hères, France*

[2] *Departamento de Física, Universidade Federal de Minas Gerais, ICEx, Caixa Postal 702, 30123-970 Belo Horizonte, Minas Gerais, Brazil*

[3] *European Synchrotron Radiation Facility, Boîte Postale 220, 38043 Grenoble Cedex, France*

Small angle neutron and x-ray scattering methods are used to investigate the structure of dilute suspensions of two different ferrofluid systems dispersed in soft polyacrylamide hydrogels. It is found that the particles in the fluid are fractal aggregates composed of smaller particles of radius ca. 5 nm. The fractal dimension is strongly dependent on sample, taking the value 1.7 in the first sample and 2.9 in the second sample. In the presence of a magnetic field the aggregates orient, but are restricted in both their translational and rotational freedom. The effect of the gel elasticity is treated as a hindrance to the orientation process.

## I. INTRODUCTION

Because of their striking physical properties and numerous practical applications, ferrofluids have been the subject of many investigations over the past two decades [1]. While the surface and near-surface structures of these liquids in the presence of a magnetic field can be detected straightforwardly using optical methods, their strong absorption of light in the visible region makes it necessary to use either thin films or small angle neutron or x-ray scattering (SANS or SAXS) for investigations of their internal structure [2–6]. More recently, x-ray photon correlation spectroscopy has been used to investigate the dynamics of one of these systems in the presence of a magnetic field [7]. Some applications of these fluids involve dispersion inside a soft rubber: introduction of a magnetic field produces a macroscopic deformation of the rubber, which can then be used as a mechanical actuator in various devices [8]. This paper reports scattering measurements using SANS and SAXS on ferrofluids dispersed in soft gels. The systems investigated in the present paper are dilute, i.e., the distance between magnetic particles is much greater than their size. This condition yields gels that are sufficiently transparent for light scattering experiments to be performed on samples of macroscopic size. To investigate the properties of these systems at a finer spatial scale, however, SANS or SAXS techniques become necessary.

The aim of the present measurements is to determine the properties of the aggregates present in these suspensions and to study the effect of an applied magnetic field when the particles are trapped in the elastic medium of a soft polymer gel. This change in the environment of the ferrofluid also implies investigating the effects of the surrounding polymer matrix upon the magnetic particles.

*Actual address: Departamento de Física - CCE, Universidade Federal de Viçosa, 36570-900, Viçosa/MG, Brazil.
Electronic address: alvaro@ufv.br

## II. EXPERIMENTAL SECTION

The polyacrylamide gels were prepared from aqueous solutions of acrylamide (Acros) into which was mixed 1/30 by weight of N-N′ methylene bisacrylamide. Polymerization was carried out at 50 °C with ammonium persulfate and TEMED according to standard recipes [9].

Two samples of ferrofluid were used in these experiments, purchased from two different sources: sample 1 was ferrofluid EMG-408 from Ferrotec, USA; sample 2 was M300 from Sigma Hi Chemical, Japan. The original concentration of ferrofluid EMG-408 was 1.1% by volume, with a saturation magnetization of 6 mT; that of the M300 was 11% by volume, having saturation magnetization 32 mT. Both samples are composed of magnetite and stabilized with a surfactant layer (sodium dodecyl-benzene sulfate for the M300 samples). No information on the surfactant in the EMG-408 samples was supplied by the manufacturer. When the measurements reported here were undertaken, sample 1 had outlived its recommended shelf life by two years. Sample 2 was freshly purchased. These samples were diluted and measured both in the liquid state and suspended inside polyacrylamide gels.

Ferrofluid from sample 1 was dispersed into the gel precursor liquid at two concentrations to yield 2.5% polyacrylamide gels containing respectively, $2.6\times10^{-4}$ g/cm$^3$ and $2.6\times10^{-3}$ g/cm$^3$ magnetic particles. Ferrofluid sample 2 was dispersed into 2.5% polyacrylamide gels containing respectively, $3.5\times10^{-4}$ g/cm$^3$ and $3.5\times10^{-3}$ g/cm$^3$ magnetic particles (sample 2a). A further specimen of sample 2 (designated sample 2b) was prepared at the higher of these two concentrations in a polyacrylamide gel of polymer concentration 4%.

A notable feature of these samples containing ferrofluids is that, unlike identical gels in the absence of the ferrofluid, they exhibit syneresis upon gelation. The release of surfactant into the polymer generates a ternary system that reduces the solvent quality in the polymer/surfactant concentration range explored. The systems described here underwent a contraction of about 15%. The syneresis, however, became visible only several weeks after the completion of the measurements. In this process, the supernatant liquid displays the characteristic opalescence of surfactant micelles, while the upper surface of the gel became coated with a thin deposit of precipitated ferrofluid, a sign that the ferrofluid particles that diffuse out of the gel are stripped of their surfactant.

Small angle neutron scattering measurements were made on the D11 instrument at the Institut Laue-Langevin, Grenoble, France, working with an incident wavelength of 8 Å at three sample-detector distances, 1.1 m, 3 m, and 10 m. In these measurements the ferrofluid was suspended in a polyacrylamide gel prepared in an $H_2O/D_2O$ mixture (45/55 by volume), at which the signal from the polyacrylamide vanishes. Exposure times were between 30 min and 1 h. Corrections for incoherent scattering were applied using the signal from an identically prepared gel sample containing no ferrofluid, following the subtraction procedure described in Ref. [10].

The SAXS measurements were made on the small angle scattering instrument on beamline BM2 at the ESRF, Grenoble. Two incident energies were used, 16 keV and 7.9 keV, with sample to detector distances of 2.10 m and 0.32 m. The detector was a two-dimensional CCD array (Princeton), cooled by a Peltier effect device and equipped with a phosphor screen. A demagnifying optical fiber bundle is used as a light guide from the phosphor screen to the CCD device. Measurements were made at room temperature (ca. 20 °C), with typical exposure times of 100 s. The blank sample used for background subtraction was a polyacrylamide gel of the same composition, not containing ferrofluid.

For the SAXS measurements in a magnetic field, the samples were prepared in 1.5 mm Lindemann capillaries and placed between the poles of a permanent magnet, the gap of which can be continuously varied between 24 cm and 3 mm, giving a maximum field strength of 1.6 T. The elements of this magnet can be arranged so that the magnetic field is either parallel or perpendicular to the incident beam [11].

## III. RESULTS AND DISCUSSION

### A. Zero field behavior

The results of the SANS measurements, shown in Fig. 1(a), display two different characteristic scattering regions each having approximately linear behavior in a double logarithmic representation. The two curves have identical shapes, but the data from the more concentrated sample is one order of magnitude more intense: the intensity is thus proportional to the concentration, as expected in the dilute regime. The low-$q$ region has a slope of approximately $-1.7$; this behavior is characteristic of aggregates builtup from smaller units by diffusion limited cluster aggregation [12,13]. In the high-$q$ region the slope is close to $-4$, characteristic of Porod scattering from the smooth surfaces of the elementary particles. In these spectra, the signal from the surrounding polyacrylamide matrix is absent, owing to the contrast match between solvent and polymer. These measurements allow the

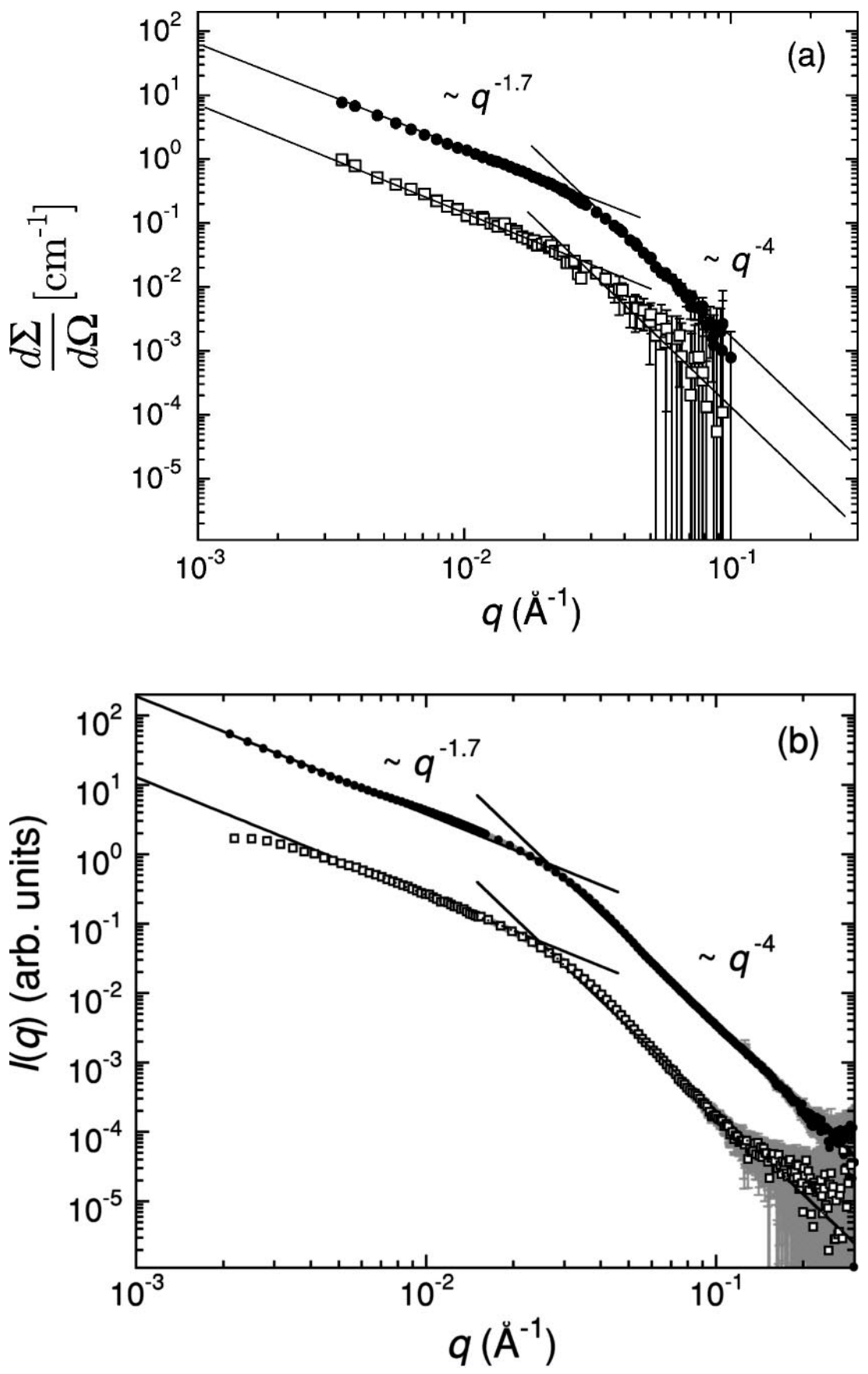

FIG. 1. (a) Neutron and (b) x-ray scattering curves from sample 1 at two ferrofluid concentrations: 0.26 mg/cm$^3$ (□) and 2.6 mg/cm$^3$ (●). Since in each technique the observations were made under identical conditions, the intensities are proportional to the concentration of ferrofluid in the sample.

specific surface area of the magnetite aggregates to be calculated, using the Porod relationship [14,15]

$$\frac{\Sigma}{V} = \frac{\pi(1-\phi)\lim_{q\to\infty} I(q)q^4}{\int q^2 I(q)dq}, \tag{1}$$

in which the integral is taken from 0 to $\infty$ and $\phi$ is the volume fraction of magnetite particles in the gel. Note that Eq. (1) defines the specific surface area *per unit volume of magnetite*. The more commonly employed expression containing an extra factor of $\phi$ in the numerator [15], corresponds to the specific surface area per volume of sample. In the present case the upper limit of the integral was taken as the extrapolation to infinity of the Porod $q^{-4}$ behavior. Owing to the high degree of dilution, the factor $(1-\phi)$ can be taken as unity. Using the assumption that the elementary particles are spheres of radius $r_0$, we have

$$\frac{\Sigma}{V} = \frac{3}{r_0}. \tag{2}$$

From the data of Fig. 1(a)

TABLE I. Specific surface area of ferrofluid particles, calculated from Eq. (1).

| | Sample | Gel conc. ($g\ cm^{-3}$) | Ferrofluid conc. ($g\ cm^{-3}$) | $\frac{\Sigma_f}{V}$ ($nm^{-1}$) | $r_0$ (nm) |
|---|---|---|---|---|---|
| SANS | Sample 1 | 0.025 | $2.6\times10^{-4}$ | $0.52\pm0.05$ | $5.8\pm0.5$ |
| | Sample 1 | 0.025 | $2.6\times10^{-3}$ | $0.55\pm0.01$ | $5.5\pm0.1$ |
| SAXS | Sample 1 | 0.025 | $2.6\times10^{-4}$ | $0.44\pm0.01$ | $6.9\pm0.1$ |
| | Sample 1 | 0.025 | $2.6\times10^{-3}$ | $0.45\pm0.01$ | $6.6\pm0.1$ |
| | Sample 2a | 0.025 | $3.5\times10^{-4}$ | $0.66\pm0.03$ | $4.6\pm0.2$ |
| | Sample 2a | 0.025 | $3.5\times10^{-3}$ | $0.64\pm0.03$ | $4.7\pm0.2$ |
| | Sample 2b | 0.040 | $3.5\times10^{-3}$ | $0.66\pm0.03$ | $4.6\pm0.2$ |

$$r_0 \approx 5 \text{ nm}. \tag{3}$$

Figure 1(b) shows the scattering results from SAXS measurements on the same system. The signal to noise ratio in this case is higher for three reasons. First, the electron density contrast in SAXS between ferrofluid and the polymer matrix is strongly favorable; second, the matching condition in the SANS experiment reduces the contrast between ferrofluid and surroundings; third, the incident flux of x-ray photons is greater than that of neutrons. In the high-$q$ region the SAXS spectra display the same Porod scattering feature as in the SANS data, thus showing that the signal from the polymer gel does not contribute significantly. As can be seen from the results listed in Table I, the numerical values found for $r_0$ from SAXS are close to those found from SANS, with, however, a slight excess in the SAXS value over that obtained by SANS. This discrepancy may be due in part to the scattering contribution from the surfactant layer surrounding the particles, which is expected to be masked in the contrast matched condition of the SANS observation. The similarity in shape of Figs. 1(a) and 1(b) show that the two scattering techniques yield essentially the same results in the $q$-range explored. It can be seen, moreover, that a shoulder tending towards plateau behavior becomes discernable in the lowest-$q$ region in the SAXS response of the more dilute sample (the absence of an observable Guinier region in the more concentrated sample is probably due to the stripping of the surfactant layer, which will cause larger aggregates to develop in the more concentrated sample.) This $q$ region was not reached in the SANS measurements. Fitting these data to a Guinier plot yields for the radius of gyration $R_G = 100$ nm. It follows that the aggregation number in this ferrofluid, given by $n=(R_G/r_{G0})^{1.7}$, is about 250, where $r_{G0}$ $[=(3/5)^{1/2} r_0]$ is the radius of gyration of the primary particles, assumed to be uniform spheres of radius $r_0$.

For sample 2a (also dispersed in 2.5% gel and with ferrofluid concentrations shown in Table I), the SAXS results are displayed in Fig. 2, where the spectrum for the ferrofluid diluted in water is shown together with that for the same sample suspended in a polyacrylamide gel. In neither of these responses is an extended fractal regime clearly apparent, unlike the case of sample 1. It is also apparent that the two curves, in solution and in the gel, have different shapes. To analyze these results in a consistent manner a model is required. To this end we adopt the expression of Chen and Teixeira [16,17], which describes the structure factor $S(q)$ of fractal aggregates composed of elementary particles of radius $r_0$,

$$S(q) = 1 + \frac{1}{(qr_0)^D} \frac{D\Gamma(D-1)}{[1+(q\xi)^{-2}]^{(D-1)/2}} \times \sin[(D-1)\arctan(q\xi)]. \tag{4}$$

Here $\Gamma(x)$ is the gamma function, $D$ is the fractal dimension of the aggregates, and $\xi$ is the upper cutoff size in the distribution. The radius of gyration is related to the cutoff size by

$$R_G^2 = D(D+1)\frac{\xi^2}{2}. \tag{5}$$

To obtain the total scattering function, the structure factor $S(q)$ is multiplied by the form factor of a sphere,

$$P(q) = \left[3\frac{\sin(qr_0)-(qr_0)\cos(qr_0)}{(qr_0)^3}\right]^2. \tag{6}$$

In the fitting procedure to the above expression, a distribution of elementary particle sizes is required in order to damp the oscillations that would arise from the form factor of the elementary spheres if their radius $r_0$ were monodisperse. For this purpose a lognormal distribution was assumed having the form

$$p(r)=\frac{1}{\sqrt{2\pi}\beta r_0}\exp\left[-\frac{[\ln(r/r_0)]^2}{2\beta^2}\right], \tag{7}$$

in which the characteristic width of the distribution (polydispersity factor) is taken to be $\beta=0.25$. Nonlinear least squares fitting of the resulting scattering function to the data from sample 2 (continuous curves in Fig. 2) yield values of $D$ between 2.8 and 2.95, indicating that the aggregate structure in this sample is much more compact than sample 1. In this fitting procedure, the parameters $D$, $\xi$, and $r_0$ are taken as free variables. The numerical results, listed in Table II, are in good agreement with those found directly from the final slope using Eq. (1), while the fits of Eq. (4) are in acceptable agreement with the data.

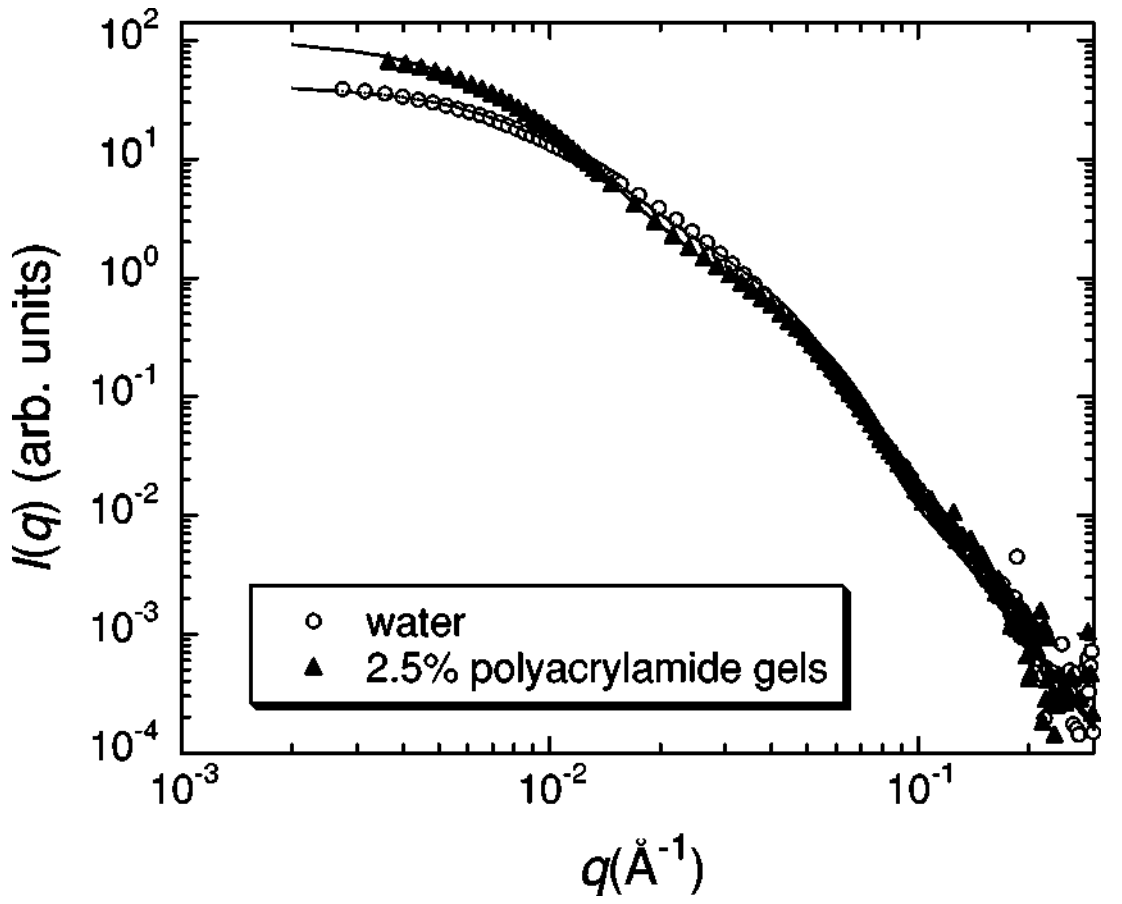

FIG. 2. X-ray scattering curves from sample 2a at the higher ferrofluid concentration and the corresponding ferrofluid solution. The continuous curves are the fits to Eq. (4) multiplied by the form factor.

TABLE II. Parameters of fit to Eq. (4) of SAXS curves for ferrofluid M300 in water and in gel.

| Medium | $D$ | $\xi$ (nm) | $R_G$ (nm) | $r_0$ (nm) |
|---|---|---|---|---|
| Water | 2.8 | 9.5 | 22 | 6.7 |
| 2.5% polyacrylamide gel | 2.95 | 12 | 29 | 5.1 |

In Table II it can be seen that the cluster size $\xi$ is larger in the gel than in solution, while the reverse is true for the elementary particle radius $r_0$. This behavior can be understood in terms of the surfactant layer around the particles that ensures steric repulsion. During polymerization of the gel some of this surfactant is removed by the polymer, thus favoring partial coalescence of the clusters, with a concomitant increase in their size [18,19]. At the same time, removal of the surfactant reduces the apparent radius of the primary particles seen by SAXS, as well as causing syneresis in the gel.

### B. Behavior in a magnetic field

When an unconfined ferrofluid is placed in a magnetic field, the aggregates group themselves into strings stretched along the direction of the magnetic field [20]. The scattering pattern from this arrangement consists of an intense band directed at right angles to the field. The same pattern is also seen to develop when the ferrofluid is mixed into a gel precursor fluid and placed in a magnetic field. After gelation has taken place, the resulting scattering pattern remains permanently anisotropic, even when the sample is removed from the field. When, however, a gel containing a dilute ferrofluid is prepared in the absence of magnetic field and is then placed in an external magnetic field **B**, it is expected that the aggregates will tend to orient along the field, but that they will not be able to migrate to form the same superstructures as in the free liquid. Furthermore, in the gel matrix, alignment of the clusters will be also hindered by elastic constraints. This implies that magnetization of this system requires higher fields, or in other words, that the magnetic susceptibility will be reduced by the gel elasticity.

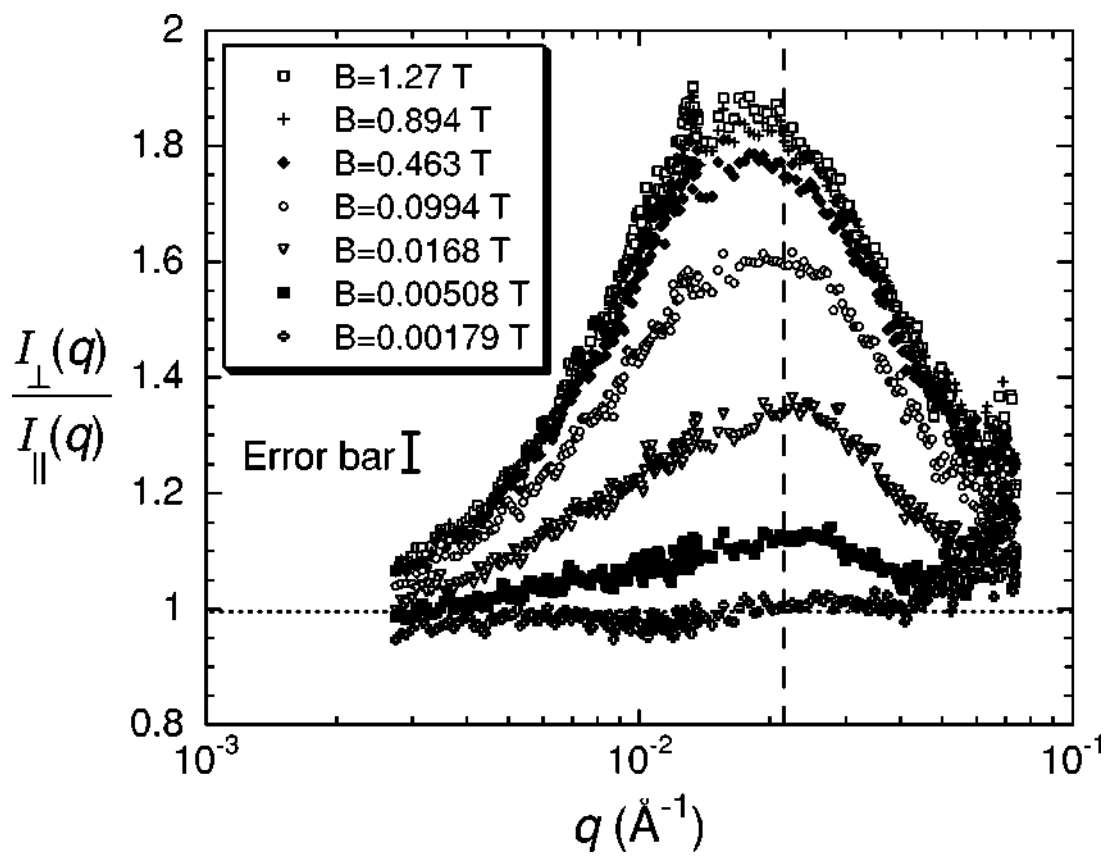

FIG. 4. Eccentricity of SAXS images for sample 2a. Ratio of intensities $I_\perp/I_\parallel$ as a function of wave vector $q$. The vertical line indicates the scattering vector $q=2\pi/R_G$.

In Fig. 3 two-dimensional SAXS patterns from a gel containing ferrofluid sample 2 are shown at different values of **B** between zero and 1.2 T. When the applied field is in the horizontal direction, a scattering pattern is generated that is elongated vertically. To a good approximation the shape of the isointensity curves is elliptical with approximately the same value of aspect ratio $\delta=1.2$ throughout the region $0.003$ Å$^{-1}<q<0.015$ Å$^{-1}$. Figure 4 shows the ratio of the intensities in the perpendicular ($I_\perp$) and parallel ($I_\parallel$) directions as a function of wave vector $q$. It can be seen that $I_\perp(q)/I_\parallel(q)$ displays a broad maximum whose peak is located just below $q=2\pi/R_G$ (vertical line), where the largest dimension of the aggregates is the diameter, $2R_G\approx 58$ nm. At the lowest values of $q$ the scattering becomes isotropic in the thermodynamic limit, while at higher-$q$ isotropy is recovered because the primary particles in the aggregates are spherical. Such behavior suggests that the aggregates are nonspherical assemblies of primary particles that become oriented in the magnetic field: as **B** increases the radius of gyration in the parallel direction increases, while that in the perpendicular direction decreases. Under these conditions, in the intermediate $q$ range, the intensity scattered in the per-pendicular direction exceeds that in the parallel direction. At high $q$, the scattering comes from the surface of the primary particles, which are isotropic. Figure 5 shows the regrouped data from these figures, where the radial average is taken over a 10° angular sector as a function of **B**. The arrow in this figure indicates the value of $q=0.005\,06$ Å$^{-1}$ that was selected to study the effect of the magnetic field on the scattering response.

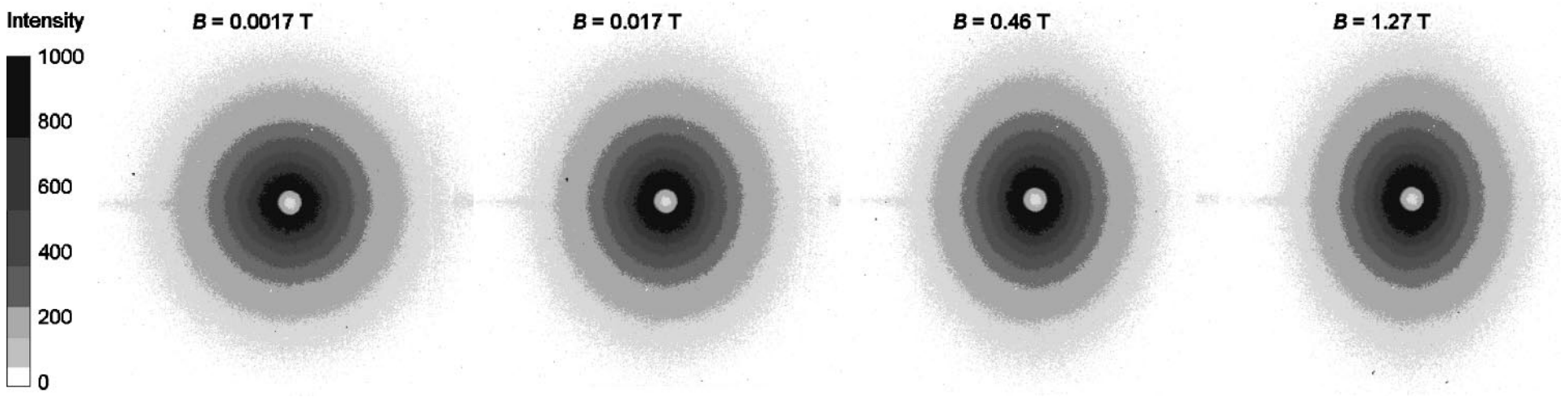

FIG. 3. SAXS patterns from sample 2a.

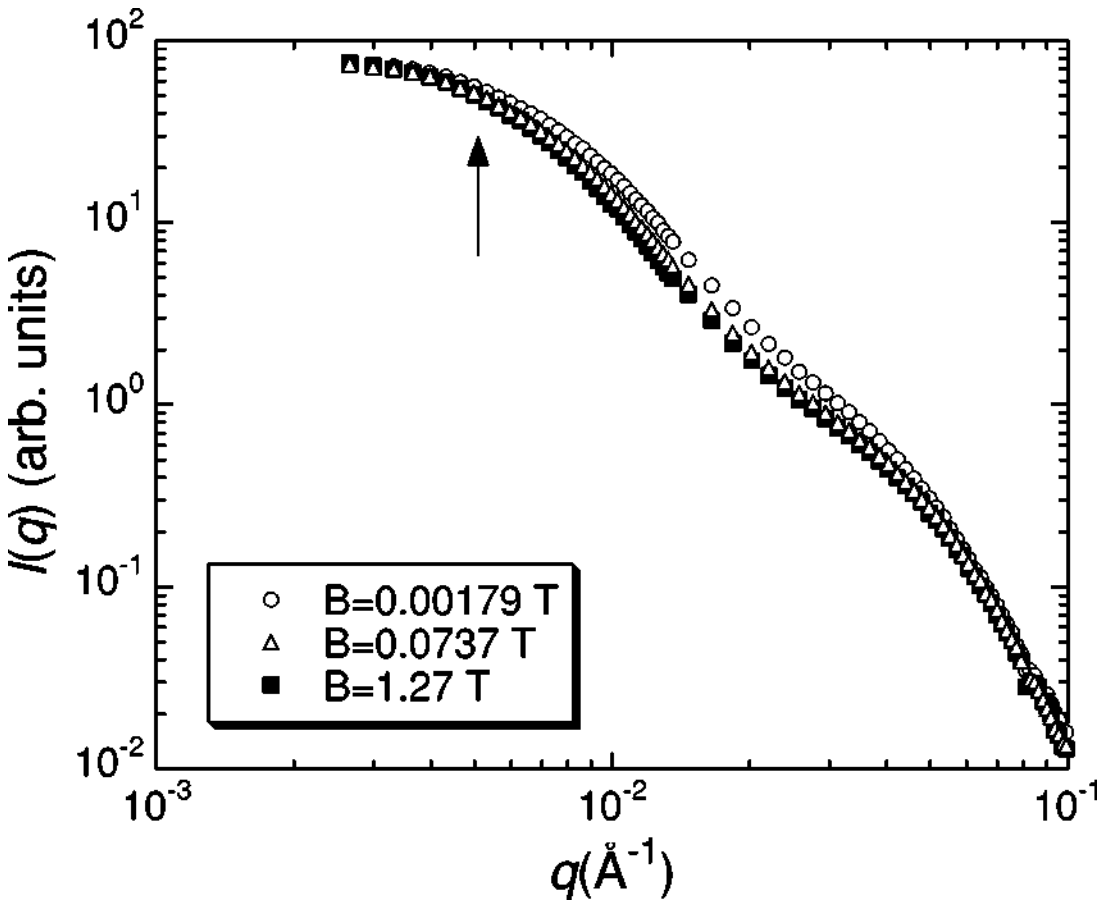

FIG. 5. Scattered intensity in the axis parallel to the magnetic field: the intensity decreases with **B**.

To analyze the shapes of the curves, we first note that the scattered intensity must be an even function of $q$: this is illustrated in the Guinier approximation, according to which the intensity varies as

$$I(q)=I(0)\exp[-(qR_G)^2/3], \tag{8}$$

where $R_G$ is the mean radius of gyration of the clusters and $I(0)$ is the scattering intensity at zero wave vector. This symmetry of $I(q)$ about the radius of gyration requires that the scattering pattern be independent of the sign of **B**, and hence the magnetic field dependence of $I(q)$ must also be an even function of **B**.

Figure 6 shows the effect of magnetic field on the relative variation of the scattered intensity for sample 2 in the polymer gels of concentration 2.5% and 4.0% (samples 2a and 2b).

$\frac{I_0(q)-I_B(q)}{I_0(q)}$: in the direction parallel to the applied field

$\frac{I_B(q)-I_0(q)}{I_0(q)}$: in the direction perpendicular to the applied field,

both measured at $q=0.005\,06$ Å$^{-1}$, i.e., in the Guinier region. Here $I_0(q)$ and $I_B(q)$ are the intensities without and with the magnetic field. At intermediate $q$, the intensity in the parallel direction decreases with increasing **B** ($\mathbf{q}\|\mathbf{B}$) and increases in the perpendicular direction ($\mathbf{q}\perp\mathbf{B}$). According to the model suggested above, we therefore assume that the elementary particles form rigid aggregates in the shape of prolate ellipsoids with principal axes of length $2a$, $2a$, and $2\delta a$. [Such a model is qualitatively consistent with the shift to lower values of $q$ in the position of the maximum of $I_\perp(q)/I_\|(q)$ with increasing **B** that is visible in Fig. 4.] We also assume that the net magnetic dipole moment is oriented along the major axis. The scattering from a free aggregate in the Guinier region [$qR_G<1$ in Eq. (8)], according to Eqs. (8), (A3), and (A4), is then given by

$$\frac{|I_B(q)-I_0(q)|}{I_0(q)}=q^2A\left[1-3\left(\frac{k_BT}{mB}\right)\mathcal{L}\left(\frac{mB}{k_BT}\right)\right], \tag{9}$$

in which $\mathcal{L}(x)$ is the Langevin function, $m$ is the total magnetization of the aggregate and $k_B$ is the Boltzmann constant (see the Appendix). Equation (9) is the expected form for the square of the magnetization of a free magnetic dipole [21]. The parameter $A$ depends on the size $a$ and eccentricity $\delta$ of the aggregate. In the parallel direction,

$$A=\frac{2a^2(\delta^2-1)}{15}. \tag{10}$$

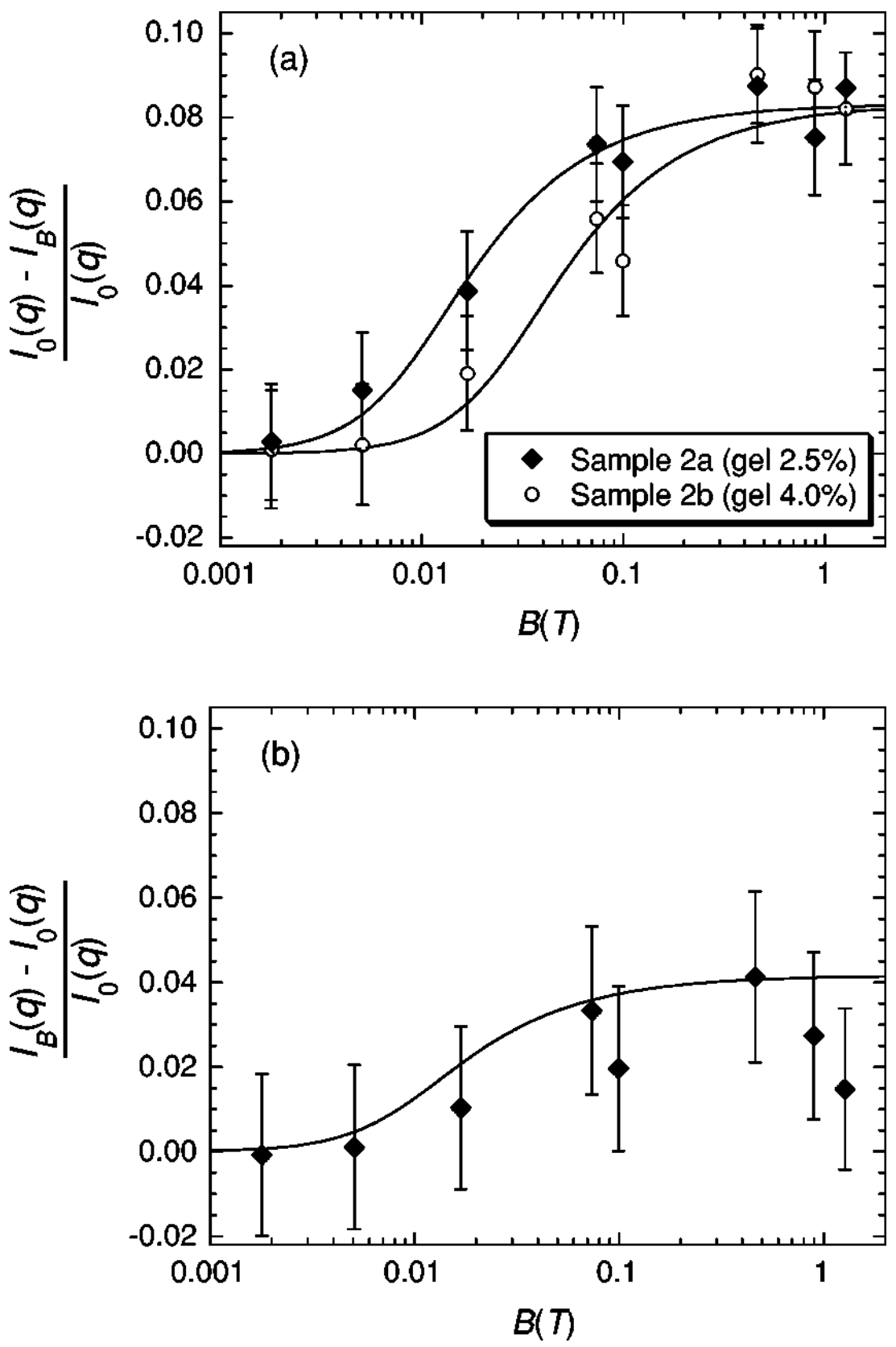

FIG. 6. Relative variation of the scattered intensity in the directions: (a) $\mathbf{q}\|\mathbf{B}$; (b) $\mathbf{q}\perp\mathbf{B}$ for samples 2a and 2b.

The analytical calculation of corrections to Eq. (9) due to the elastic constraints of the gel is quite involved and, here, for simplicity, we adopt the free particle equation as an empirical fitting function. An effective reduced magnetization $m_{eff}$ is introduced to include elastic effects. Equation (9) is fitted by a nonlinear least squares procedure [continuous line in Fig. 6(a)]. The ratio of the parameter $A$ in the parallel and perpendicular direction is 2, since the variation of the square of the radius of gyration in the parallel direction is twice that in the perpendicular direction, to conserve the volume. The results of the fits are listed in Table III. The values of $m_{eff}$ are expressed in units of Bohr magneton ($\mu_B = 9.274 \times 10^{-24}$ J T$^{-1}$).

TABLE III. Parameters of fit to Eq. (9) of variation of scattering intensity.

| Medium | $m_{eff}(\times 10^5\ \mu_B)$ | $A$ |
|---|---|---|
| 2.5% gel | $1.2\pm 0.2$ | $0.083\pm 0.003$ |
| 4.0% gel | $0.43\pm 0.08$ | 0.083 (fixed) |

On setting the eccentricity equal to $\delta = 1.2$, the value of the parameter $A$ yields an ellipsoid having $a = 24$ nm and a radius of gyration $(= a(\delta^2 + 2)^{1/2}/5^{1/2})$ of $20\pm 6$ nm, which is close to the value found in Table II. Figure 6(b) shows the equivalent variation of the intensity in the perpendicular direction and the continuous line is the theoretical curve using the same value of $m_{eff}$ as already found, with half of the amplitude in Fig. 6(a).

From the known magnetization of magnetite ($4.46 \times 10^5$ J T$^{-1}$) and the volume of the elementary particles ($r_0 = 5$ nm), it follows that their magnetic moment is $0.25 \times 10^5 \mu_B$. Moreover, the aggregation number for sample 2, given by $n = (R_G / r_{G0})^D$, where $r_{G0}$ is the radius of gyration of a sphere with radius of 5 nm, is found from Table II to be approximately 250. With this estimate an evaluation can be made of the magnetic moment of the aggregates. On the assumption that the mutual orientation of the primary particles is random, i.e., proportional to $n^{1/2}$, we obtain for the magnetic moment of the aggregates,

$$m = 4.0 \times 10^5 \mu_B. \tag{11}$$

It should be pointed out that while the elastic modulus $G$ of the polymer network acts to hinder the orientation, it will have no effect on the asymptotic limit. The value found in Eq. (11) is an estimated upper bound and should correspond to the limit of zero gel concentration.

The difference between samples 2a and 2b in Fig. 6 shows that the elasticity of the gel does indeed exert an appreciable effect on the magnetic response, since the elastic modulus $G$ is higher in the more concentrated gel. However, from a knowledge of the value of $G$ in these gels a simple estimate can be made of this effect. We consider an aggregate of radius $R_G$, embedded in an infinite elastic medium of modulus $G$, rotating through an angle $\theta$. The gel at its boundary is displaced by a distance proportional to $R_G\theta$. Since the surface area of the displaced gel is defined by $R_G^2$, the force required to produce this displacement is proportional to $GR_G^2\theta$. It follows that the energy required to rotate the aggregate through a significant angle (e.g., 1 radian) is given, to within a factor of order unity, by $E_e \approx G \times R_G^3$. This quantity may be considered as a typical potential that must be overcome in order to saturate the sample magnetization. A saturation field can then be estimated from

$$E_G = mB = G \times R_G^3. \tag{12}$$

By empirically fitting the Langevin equation to the field response, elastic effects are incorporated as a diminished effective free dipole moment. In this case, the saturation field is estimated from

$$E_T = m_{eff} B = k_B T. \tag{13}$$

Equating saturation fields from the above estimations allows one to obtain the relation between the effective moment and gel elastic modulus. In the limit of strong elasticity one would then expect

$$m_{eff} \propto G^{-1}. \tag{14}$$

In Ref. [22] it was found that for a 4% polyacrylamide gel similar to that used here, the value of $G$ is $\approx 7$ kPa, while for the 2.5% sample, it is about 2 kPa. These values of $G$ yield $B = 0.05$ T and $B = 0.015$ T for the 4% and 2.5% gel, respectively, in reasonable agreement with the observed points of inflection in the two curves in Fig. 6(a) and also with the values reported in Table III, according to Eq. (14).

## IV. CONCLUSIONS

Small angle neutron and x-ray scattering measurements on dilute ferrofluid suspensions in polyacrylamide gels show that the particles have a fractal character, the dimensionality of which depends on the history of the ferrofluid under investigation. The scattering curves of the fresh samples are found to be well described by the Chen-Teixeira expression for scattering by fractal aggregates. Comparison with the ferrofluid in liquid suspension shows that during the polymerization process in the gel, a further tendency of the aggregates to coagulate is observed. This is explained by the stripping of the protective layer of surfactant from the magnetic particles by the interaction with the polymer, thus reducing their mutual repulsion during the period that the system is still in the sol state.

When an external magnetic field is applied to these ferrogels, the aggregates, which are constrained in both their translational and rotational degrees of freedom, tend to align by rotating along the direction of the magnetic field. The aggregate anisotropy can be thus deduced. An effective magnetization $m_{eff}$ is introduced to take into account the elastic effects of the surrounding gel. The values of $m_{eff}$ calculated from the fit to a Langevin type equation are in acceptable agreement with that calculated from the primary particle radius and the aggregation number, obtained from the wave vector dependence of the scattering intensity. Furthermore, the estimated elastic contribution to $m_{eff}$ is in agreement with the observed difference between samples of different polymer concentration.

## ACKNOWLEDGMENTS

We are grateful to the Institut Laue-Langevin, Grenoble, for access to the D11 instrument and to the European Synchrotron Radiation Facility, Grenoble, for beam time on the BM2 beam line. We particularly wish to thank J. Chavanne of the ESRF for designing and supplying the permanent magnet arrays used in the variable magnetic field device. A.V.N.C.T. expresses his gratitude to CAPES, Brazil, for a study grant.

## APPENDIX: MEAN RADIUS OF GYRATION OF AN ELLIPSOID

An ellipsoid of revolution is characterized by two major axes $2a$ and $2\delta a$, where $\delta$ is the eccentricity of the ellipsoid. In some specific direction $z$, the mean square radius of gyration is calculated considering all possible orientations with probability density distribution given by the Boltzmann distribution.

If $\theta$ is the angle between the major axis of the ellipsoid and the $z$ axis, it is possible to show that the mean square radius of gyration of an ellipsoid with mass $M$ is related to its major inertia moments $I_{xx}$, $I_{yy}$, and $I_{zz}$ by

$$R_{Gz}^2=\frac{[I_{xx}+I_{yy}-I_{zz}]}{2M}=\frac{a^2}{5}[(\delta^2-1)\cos^2\theta+1]. \quad \text{(A1)}$$

If the ellipsoid has a permanent magnetic dipole moment parallel to the major axis, the interaction energy between the dipole and the magnetic field is $-mB\cos\theta$ and, with Eq. (A1) is

$$\langle R_{Gz}^2\rangle=\frac{a^2}{5}+\frac{a^2}{5}(\delta^2-1)\frac{\int_{-\alpha}^{\alpha}dx\,x^2e^x}{\alpha^2\int_{-\alpha}^{\alpha}dx\,e^x}, \quad \text{(A2)}$$

where $\alpha=mB/k_BT$. The ratio of the integral terms is the mean square value of a free magnetic dipole in a homogeneous magnetic field. It is straightforward to show that

$$\langle R_{Gz}^2\rangle=\langle R_{Gz}^2\rangle_0+A\left[1-3\left(\frac{k_BT}{mB}\right)\mathcal{L}\left(\frac{mB}{k_BT}\right)\right], \quad \text{(A3)}$$

where $\langle R_{Gz}^2\rangle_0=a^2(\delta^2+1)/15$ is the mean square radius of gyration in the $z$ axis in the absence of $\mathbf{B}$ and $A=2a^2(\delta^2-1)/15$. Repeating the same calculation in some direction perpendicular to the magnetic field (for example, the $x$ axis) we have

$$\langle R_{Gx}^2\rangle=\langle R_{Gx}^2\rangle_0-\frac{A}{2}\left[1-3\left(\frac{k_BT}{mB}\right)\mathcal{L}\left(\frac{mB}{k_BT}\right)\right], \quad \text{(A4)}$$

which is essentially the same form of Eq. (A3) except that the radius of gyration diminishes with $B$ and the multiplying factor is one half that of Eq. (A3). The zero-field radius of gyration is the same in all directions.